\documentclass[aps,twocolumn,amsmath,floatfix]{revtex4-1}
\usepackage{amssymb}
\usepackage{graphicx}
\usepackage{array}
\usepackage{hhline}
\usepackage{longtable}
\usepackage{bm}
\usepackage{hyperref}
\usepackage{color, soul}
\usepackage{wasysym}

\begin{document}

\newcommand{\rsp}[1]{\hspace{-0.15em}#1\hspace{-0.15em}}

\title{Dynamics of evaporative colloidal patterning}


\author{C. Nadir Kaplan$^1$, Ning Wu$^2$, Shreyas Mandre$^3$, Joanna Aizenberg$^{1, 4, 5, 6}$, and L. Mahadevan$^{1, 4, 5, 7}$}
\affiliation{
\\$^{1}$School of Engineering and Applied Sciences, Harvard University, Cambridge, MA 02138, USA.
\\$^{2}$Department of Chemical Engineering, Colorado School of Mines, Golden, CO 80401, USA.
\\$^{3}$School of Engineering, Brown University, Providence, RI 02912, USA.
\\$^{4}$Wyss Institute for Biologically Inspired Engineering, Harvard University, Boston, MA 02115, USA.
\\$^{5}$Kavli Institute for Bionano Science and Technology, Harvard University, Cambridge, MA 02138, USA.
\\$^{6}$Department of Chemistry and Chemical Biology, Harvard University, Cambridge, MA 02138, USA
\\$^{7}$Department of Physics, Harvard University, Cambridge, MA 02138, USA.}

\begin{abstract}
Drying suspensions often leave behind complex patterns of particulates, as might be seen in the coffee stains on a table. Here we consider the dynamics of periodic band or uniform solid film formation on a vertical plate suspended partially in a drying colloidal solution. Direct observations allow us to visualize the dynamics of the band and film deposition, and the transition in between when the colloidal concentration is varied. A minimal theory of the liquid meniscus motion along the plate reveals the dynamics of the banding and its transition to the filming as a function of the ratio of deposition and evaporation rates. We also provide a complementary multiphase model of colloids dissolved in the liquid, which couples the inhomogeneous evaporation at the evolving meniscus to the fluid and particulate flows and the transition from a dilute suspension to a porous plug. This allows us to determine the concentration dependence of the bandwidth and the deposition rate. Together, our findings allow for the 
control of drying-induced patterning as a function of the colloidal concentration and evaporation rate. 
\end{abstract}

\date{\today}
\maketitle


\section{Introduction}

Colloidal self organization occurs in systems such as opals~\cite{Opal2}, avian skin~\cite{Prum}, photonics~\cite{Photonics1, Photonics2, Photonics3}, and tissue engineering~\cite{Tissue1}. An approach to colloidal patterning is via evaporation-driven deposition of uniformly dispersed particles in a volatile liquid film~\cite{Hatton, Snoeijer1, Bigioni, Deegan1, Deegan2, Deegan3, Popov, Crosby, Bodiguel}. The basic mechanism of evaporation-driven patterning involves vapor leaving the suspension more easily along the liquid-air-substrate triple line (the contact line), resulting in a singular evaporative flux profile. The combination of the contact line pinning and a singular evaporative flux there generates a fluid flow that carries the dissolved particles towards the edge of the film~\cite{Deegan1, Deegan2}. The advected colloids then get arrested near the contact line to form patterns such as continuous solid films, or regular bands~\cite{Adachi, Stone2, Chang, Stone3}, that are laid down along the 
substrate. 

\begin{figure}
\centering
\includegraphics[width=1\columnwidth]{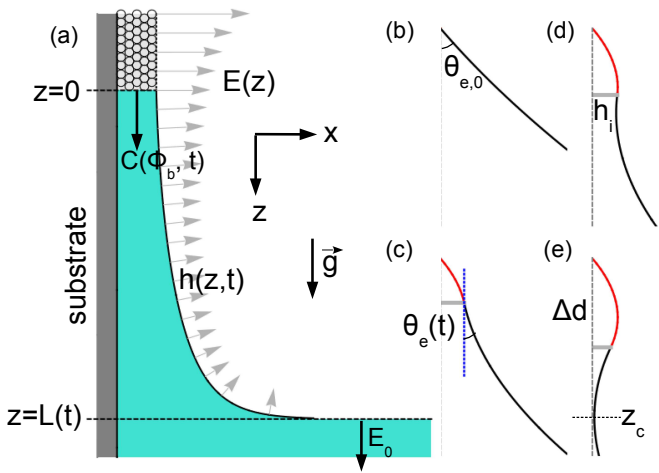}
\caption{{\bf Schematics of a drying suspension on a vertical substrate.} (a) A schematic of the geometry and variable definitions for banding on a vertical substrate. (b), (c), (d), (e) The evolution of the meniscus deformation while leaving behind a colloidal deposit, and the corresponding variable definitions. Grey full line indicates the location of the solid-liquid interface, while the red curves represent the local thickness of the solid deposit. }
\label{fig:schematics}
\end{figure}

The evaporative patterns form via a complex dynamics. The propagation rate of the interface separating the liquid and the colloidal deposit is controlled by the local particle concentration in the solution (Fig.~\ref{fig:schematics}(a)) and the velocity of the viscous capillary flow transporting the colloids. During this process, the liquid meniscus pinned to the edge of the deposit deforms as a function of the rate of particle transport and evaporation, in turn dictating the formation of either a continuous film or a periodic band (Fig.~\ref{fig:schematics}(b)--(e) and Fig.~\ref{fig:experiment0}(a)--(f)). As a result, the particles self-organize into various forms of ordered and disordered states\cite{Yodh1} as a function of the deposition speed and the local evaporation rate~\cite{Snoeijer1, Goehring}. An additional complexity is that the fluid flow regime changes dramatically over the course of the drying process. Initially, we have flow in a thin film that is characterized by the Stokes regime away from 
the deposition front where the particle concentration is low. Near the deposition front, the liquid enters a porous region that is itself created by the particulate deposits at the solid-liquid interface, leading to a Darcy regime. Early models~\cite{Adachi, Stone2, Stone3, Witten, Kaya} focused on understanding the singular evaporative flux and the related particulate flux, leaving open mechanisms for the filming-banding transition, the deposition front speed that sets the rate of patterning, and the Stokes-Darcy transition, questions we answer here.   

Here we use a combination of experimental observations and theoretical models of the interface growth and colloidal patterning to understand the dynamics of periodic banding, and its transition to the deposition of a continuous film as a function of the particle concentration. Based on our observations, we formulate two complementary theories: (1) A coarse-grained two-stage model, which consists of a hydrostatic stage until the meniscus touches down the substrate, followed by rapid contact line motion terminated by its equilibration. This minimal model allows us to explain the geometry of the periodic bands as a function of the deposition rate. (2) A detailed multiphase model of the drying, flowing suspension allows us to account for the Stokes-Darcy transition, and couples the evaporation rate, fluid flow, the meniscus height, the distribution of particle concentration, and the dynamic interface velocity as a function of the initial particle concentration. This theory leads to explicit predictions for the deposition rate, and the banding-filming transition in good quantitative agreement with the measurements.

The present study is organized as follows: Experiments are described in Sec.~II. Based on experimental evidence, we develop the minimal two-stage model in Sec.~III. This model is complemented by the multiphase model in Sec.~IV. Concluding remarks are given in the final section of the paper.

\section{Experiments}
\noindent
{\bf Methods.} Our experiments were performed by partly immersing a vertical unpatterned silicon and  glass substrates in a dilute colloidal suspension of colloidal spheres (see Fig.~\ref{fig:schematics}(a) for the experimental setup). Silicon and glass substrates were used for evaporative colloidal coating. They were first cleaned in a mixture of sulfuric acid and hydrogen peroxide (3:1) at 80~$^o$C for one hour. They were then rinsed with deionized water thoroughly and treated with oxygen plasma for 1 minute immediately before use. Colloidal particles were either synthesized (375 nm PMMA spheres) by using surfactant-free emulsion polymerization or purchased from Life Technologies (1 $\mu$m latex spheres). Before the evaporative deposition, particles were centrifuged four times and were re-dispersed in deionized water. The substrate was mounted vertically and immersed partially in a vial containing the colloidal suspension. Water was evaporated slowly over a period of $\sim$12~hr to 2~d in an oven that was placed on a vibration-free table. The solvent evaporation rate was controlled by the temperature of the oven and was measured by putting a second vial filled with colloidal suspension but without the substrate. Images of the band structures and colloidal packing were taken by both optical microscopy (Leica DMRX) and scanning electron microscopy (Zeiss Ultra). A custom-built side view microscope (Olympus BX) was also used to image the \textit{in situ} movement of meniscus and colloids during the band formation on glass substrates. 
\\

\begin{figure}
\centering
\includegraphics[width=1\columnwidth, clip=true]{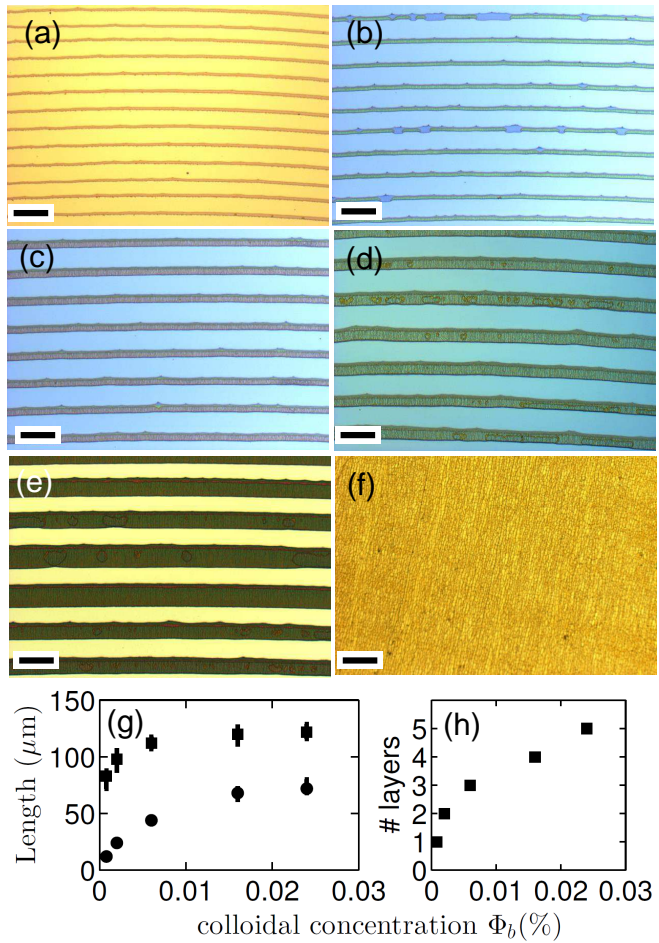}
\caption{{\bf Periodic bands and uniform films of colloidal deposits on a silicon substrate placed vertically, as a function of the colloidal volume fraction.} The volume fraction of colloids inside the suspension are; (a) $\Phi_b=8\times 10^{-6}\,,$ (b) $\Phi_b=2\times 10^{-5}\,,$ (c) $\Phi_b=6\times 10^{-5}\,,$ (d) $\Phi_b=1.6\times 10^{-4}\,,$ (e) $\Phi_b=2.4\times 10^{-4}\,,$ (f) $\Phi_b=4\times 10^{-4}$ (uniform film). In (a)--(f) the scale bars are $200$ $\mu$m. (g) Band spacing $d$ ($\blacksquare$) and bandwidth $\Delta d$ ($\CIRCLE$) are plotted with their error bars, as a function of the colloidal volume fraction. The wavelength is given by $d+\Delta d\,.$ (h) Layer number is plotted as a function of the colloidal volume fraction.}
\label{fig:experiment0}
\end{figure}

\begin{figure}
\centering
\includegraphics[width=1\columnwidth]{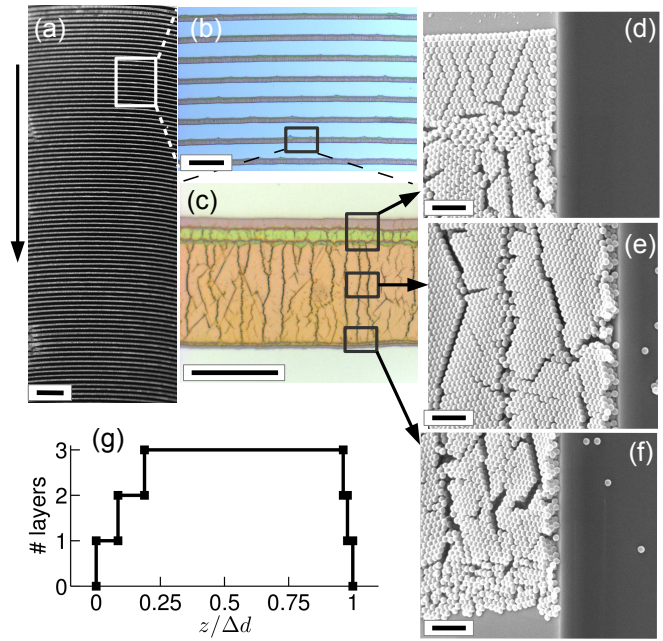}
\caption{{\bf Periodic bands in a drying dilute suspension ($\Phi_b=\mathbf{6\times 10^{-5}}$).} (a), (b) Optical micrographs show periodic bands at different magnifications ((a) scale bar: 1 mm, (b) scale bar: 200 $\mu$m). The vertical arrow in (a) indicates the direction of meniscus movement during evaporation. (c) Optical image of a single band (scale bar: 40 $\mu$m). The profile of the band is asymmetric with significant differences between the advancing (gradual transition to a thicker layer and ordered packing) and receding side (abrupt transition and random packing). (d), (e), (f) SEM images at different locations of the colloidal deposit in (c) (scale bar: 2 $\mu$m). (g) Cross-sectional profile of the band shown in (c). Position $z$ is scaled with the width of the band $\Delta d$. The meniscus moves  from left to right.}
\label{fig:experiment1}
\end{figure}

\noindent
{\bf Results and discussion.} As evaporation proceeds, two types of patterns are observed near the contact line. When the bulk volume fraction $\Phi_b$ inside the reservoir is bigger than a critical value $\Phi_c\,,$ a continuous film of particles was deposited by the receding contact line (Fig.~\ref{fig:experiment0}(f) and  Movie S1 in Supporting Information). However, when $\Phi_b<\Phi_c$, the contact line retreated leaving behind a periodic pattern of colloidal bands (Fig.~\ref{fig:experiment0}(a)--(e), Fig.~\ref{fig:experiment1} (a), (b), and Movie S2 in Supporting Information). The bands are oriented locally parallel to the receding contact line; the small curvature of each band in Fig.~\ref{fig:experiment1}(a) results from the finite size of the substrate. In Fig.~\ref{fig:experiment0}(g), we show the values of the width $\Delta d$ of a single band and the spacing between adjacent bands $d\,,$ against $\Phi_b$. 

\begin{table*}
\begin{tabular}{|c||c|c|c|c|c|c|c|}
 \hline
   & \footnotesize{Bandwidth} & \footnotesize{Spacing} & \footnotesize{Period} & \footnotesize{$\theta_e$ restoration} & \footnotesize{The deposition} & \footnotesize{Critical}\\
   & \footnotesize{$\Delta d$ ($\mu$m)}& \footnotesize{$d$ ($\mu$m)}& \footnotesize{$T$ (s)} & \footnotesize{time $T_D$ (s)} & \footnotesize{speed $C$ (m/s)} & \footnotesize{concentration $\Phi_c$} \\
   \hhline{|=#=|=|=|=|=|=|}
  \hline
  Experiment & $70$ & $100$ & $500$ & $10$ & $<E_0$ & $2.4\times 10^{-4}$\\ \hline
  Mimimal model & $100$ & $150$ & $200$ & $6$ & $0.3 E_0$ & -- \\ \hline
  Multiphase model &  $55$ & --  & $200$ & -- & $0.3 E_0$ & $6.8\times 10^{-3}$ \\ \hline 
\end{tabular}
\caption{Quantitative comparison of structural variables between the experiments and theoretical models. All values are approximate. The measurements in the experiments were obtained for the bulk colloidal volume fraction $\Phi_b\sim10^{-4}\,.$ The bulk evaporation rate in the reservoir is $E_0\sim 10^{-6}$m/s for water in atmospheric pressure and room temperature. For $\Phi_b\approx 6\times10^{-4}$, the deposition height is $H\sim 1\mu$m. The outputs of the minimal model correspond to $\epsilon=H/\ell=0.01$ (Table II) and the dimensionless deposition rate $\beta\equiv C/E_0=0.3\,.$ The results of the multiphase model are given for $\epsilon=0.01$ and $\Phi_b=2\times 10^{-3}\,.$ The wavelength is given by $d+\Delta d\,.$}
\end{table*}

High magnification optical images of a typical colloidal band (Fig.~\ref{fig:experiment1}(c)) show a range of interference colors that correspond to deposits with 1 (magenta), 2 (green), and 3 (orange) particle layers at $\Phi_b=6\times 10^{-5}\,.$ The deposits have a strongly asymmetric cross-sectional shape, as evidenced by scanning electron microscopy (SEM): as the meniscus recedes, gradual deposition of wide, well ordered colloidal layers takes place (Fig.~\ref{fig:experiment1}(d), (e)). When the band stops, a region of randomly packed particles with a sharp deposition front terminates the colloidal band left behind the moving liquid meniscus (Fig.~\ref{fig:experiment1}(f)). In this region that is only a few particles wide, the packing is disordered (Fig.~\ref{fig:experiment1}(f)). The transition from the ordered to the disordered packings is attributed to the rapidity of flow at the end of a deposition cycle, not leaving time for the colloids to anneal into an ordered structure~\cite{Snoeijer1}. Once a 
deposition cycle is complete, the resulting cross-sectional profile of the deposit at $\Phi_b=6\times 10^{-5}$ is shown in Fig.~\ref{fig:experiment1}(g). The number of maximum layers increase as a function of $\Phi_b\,,$ as demonstrated in Fig.~\ref{fig:experiment0}(h).

\begin{figure}
\centering
\includegraphics[width=1\columnwidth, clip=true]{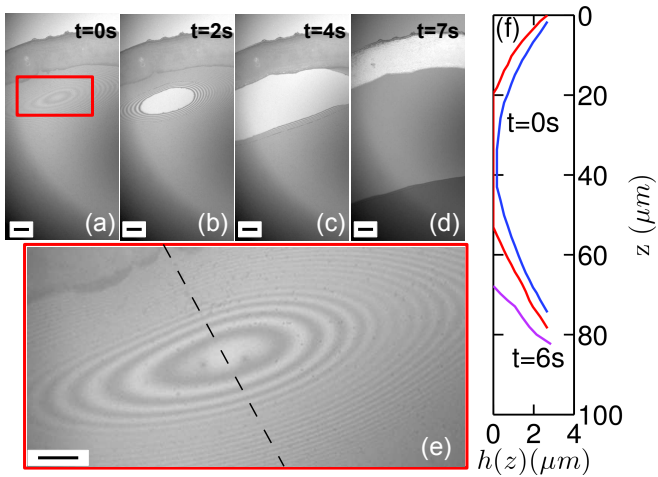}
\caption{{\bf Real-time observations of periodic banding.} (a), (b), (c), (d) A sequence of snapshots shows the meniscus break-up and the subsequent contact line motion during the formation of a new band on a vertical glass substrate in a suspension (0.06 vol~\% 1 $\mu$m latex particles in water, scale bar: 20 $\mu$m). (e) A close-up of the red box in (a) shows the interference fringes associated with the meniscus approaching the substrate (scale bar: 20 $\mu$m). (f) The meniscus profile $h(z,t)$ at $t=0\,,2\,,6$ seconds (respectively, blue, red, and magenta) extracted from the interference patterns, along the black dashed line in (e).}
\label{fig:experiment2}
\end{figure}

Light microscopy study allows us to monitor the  movement of the meniscus imposed by the evaporation. Fig.~\ref{fig:experiment2}(a)--(d), and Movie~S3 in Supporting Information show that while the deposition front advances, the meniscus approaches the substrate, as evidenced by the emerging skewed interference rings behind the contact line (Fig.~\ref{fig:experiment2}(e)). Once the meniscus touches the bare substrate and dewets, it breaks up and separates into two moving dynamic contact lines (Fig.~\ref{fig:experiment2}(b)). One of these retreats towards the just formed colloidal band (Fig.~\ref{fig:experiment2}(c)), which wicks the fluid, dries and changes its optical contrast (Fig.~\ref{fig:experiment2}(d)), while the other contact line slips until its dynamic contact angle $\theta_D$ re-equilibrates on the substrate over a time scale $T_D$ (see Table I). Colloidal particles then flow towards the stabilized contact line and start to build a new band at that location. Interference patterns allow us to measure the height of the fluid film from the substrate, as shown in Fig.~\ref{fig:experiment2}(f) in the frame of the deposit-liquid interface.

To explain the experimental results, we must account for the dynamics of meniscus deformation and break-up, the subsequent receding of the contact line, the role of particulate flow, and the transition from a dilute to a dense suspension in the vicinity of the deposition front. Thus, we first build a minimal model of the periodic banding and uniform filming, and then complement it with a multiphase approach.

\begin{figure}
\centering
\includegraphics[width=1\columnwidth, clip=true]{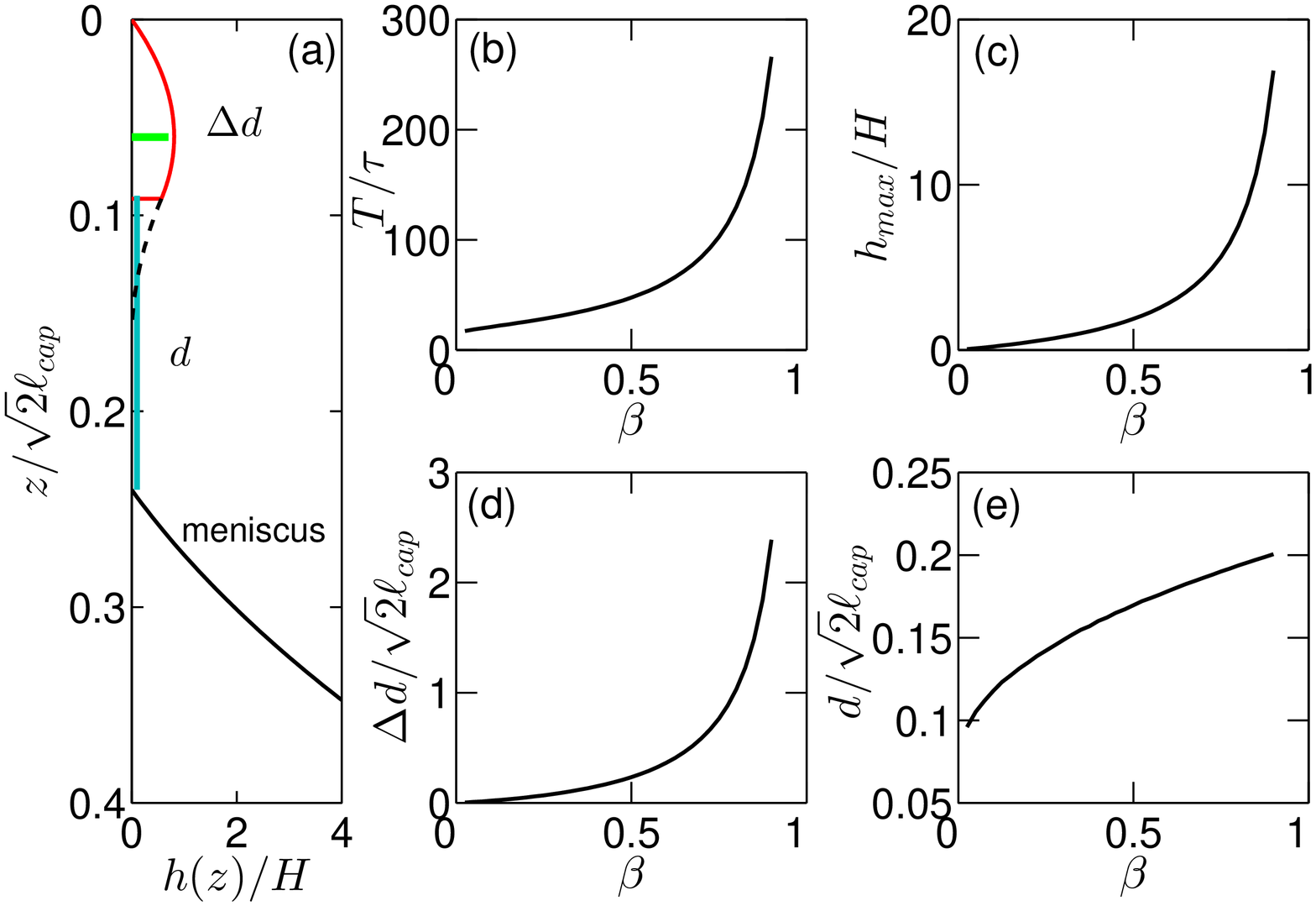}
\caption{{\bf Theoretical models of the formation of periodic bands and uniform films.}  (a) After a time period $T\,,$ a single band has formed with the width $\Delta d$ (projected length of the deposit shown by the red curve), spacing $d$ (turquoise line), and the maximum deposition height $h_{max}$ (green line). The left-over tail of the receding meniscus is demonstrated by the black dashed line ($\ell_{cap}$ is the capillary length, see Table II). The scaled dependence of (b) the period $T\,,$ and (c) the maximum band height $h_{max}$ (green line in (a)), (d) the bandwidth $\Delta d$, and (e) the spacing between bands $d$ on the dimensionless deposition front speed $\beta\equiv C/E_0\,,$ where $\tau\equiv H/E_0$ is the time scale (Table II). All curves in (a)--(e) are numerical solutions to the Eqs.~\eqref{eq:theory14}--\eqref{eq:theory15} with $\epsilon=0.01$, $\beta=0.3$, $\theta_{e, 0}=15^{o}$.}
\label{fig:theory1}
\end{figure}

\section{Minimal model for banding and filming}

To quantify the dynamics of the filming-banding transition and the deposit height as a function of the relative magnitude of evaporation and particle deposition rates, we must account for the kinetic effects associated with the meniscus deformation, break-up, and the subsequent contact line motion. Thus, we first develop a minimal hydrostatic model of the periodic band formation. This model also explains the termination of each band by a sharp deposition front, as observed in experiments. Although the meniscus break-up starts as a localized event below the maximum of the forming, convex band (see Fig.~\ref{fig:experiment2}), it then spreads laterally across the plate, resulting in a well-defined periodic banding pattern (see Fig.~\ref{fig:experiment1} (a), (b)). Therefore, we will limit ourselves to shapes and motions in a two-dimensional $x-z$ plane (Fig.~\ref{fig:schematics}).

The liquid meniscus deforms hydrostatically when the capillary forces, proportional to the surface tension $\gamma$ of the liquid-air interface, dominate over the viscous hydrodynamic forces $\mu v_f\,.$ Here $\mu$ is the dynamic viscosity of the liquid, and $v_f$ is its evaporation-induced upward flow velocity in the vicinity of the substrate. Away from the far edge of the deposit on the plate, $v_f$ is in the order of $E_0\,,$ the evaporation rate at the level of the bath. The condition for quasistatic meniscus evolution then becomes $Ca\equiv\mu E_0/\gamma\ll1\,,$ where $Ca$ is the capillary number (Table II). In this regime, the diverging evaporative flux at the contact line alters solely $v_f$ near the singularity; it has no effect on the overall meniscus evolution. Then, the hydrostatic meniscus profile $h(z, t)$ is determined by the equilibrium condition $p=p_0\,,$ where $p_0$ is the atmospheric pressure, and the pressure $p$ at the meniscus is given by
\begin{equation}
\label{eq:pressure}
p=-\gamma\kappa-\rho g \left[z-L(t)\right]+p_0\,.
\end{equation}
Here $\kappa\equiv \partial \left(\sin{\theta}\right)/\partial z$ is the curvature of the liquid-air interface, where the local angle $\theta=\theta(z)$ is defined as $\tan{\theta}\equiv \partial h/\partial z\,,$ the density of the suspension is denoted by $\rho$, and the gravitational acceleration by $g\,.$

The dynamics of the meniscus deformation is driven by two processes that act simultaneously. First, evaporation-induced flow results in the deposition of the solute near the contact line with a speed $C(\Phi_b, t)\,,$ moving the liquid-deposit wall where the meniscus is attached (Fig.~\ref{fig:schematics}(a)). Second, the level of fluid inside the container descends with a constant speed $E_0$ (Fig.~\ref{fig:schematics}(a)). Defining $L\equiv L(t)$ as the distance between the vertical level of the bath and the deposition front (Fig.~\ref{fig:schematics}(a)), which move relative to each other, $L(t)$ changes at a rate given by
\begin{equation}
\label{eq:rate}
\frac{dL}{dt}=E_0-C\,,\quad E_0\,, C>0\,.
\end{equation}
When the bulk volume fraction is smaller than the critical volume fraction ($\Phi_b<\Phi_c$), the deposition happens slower than the descent of the liquid level inside the bath ($C<E_0$) and $L(t)$ increases over time ($dL/dt>0$). As the evolving $L(t)$ changes the curvature of the concave meniscus due to the gravity when $p=p_0$ (see Eq.~\eqref{eq:pressure}), the dynamic contact angle $\theta_e(t)$ becomes smaller than the equilibrium contact angle $\theta_{e, 0}$, and decreases monotonically in time ($d\theta_e/d t<0\,,$ see Fig.~\ref{fig:schematics}(b)--(e)). Then the meniscus touches down on the substrate at a location $z_c$ behind the deposition front (Fig.~\ref{fig:schematics}(e)). This causes the meniscus to break to form two contact lines, one that moves towards the deposit, and another that recedes rapidly until it eventually re-equilibrates at a distance where the contact angle regains its equilibrium value $\theta_{e,0}$. As the process repeats, periodic bands are formed. Conversely, when $\Phi_b>\Phi_c$, $C>E_0$,  so that $dL/dt<0$. Then, $\theta_e>\theta_{e,0}$ and $d\theta_e/d t>0$, so that the meniscus moves away the plate everywhere, and a continuous deposition film will be laid out by the deposition front in this regime. The transition between the banding and filming happens at $\Phi_b=\Phi_c$ where the evaporation and deposition speeds are matched, \textit{i.e.} $E_0=C(\Phi_c) $. In this case the size of the domain will always be equal to its equilibrium value $L=L_0\,.$  

In the frame comoving  with the deposit-liquid wall at a dimensionless deposition speed $\beta\equiv C/E_0\,,$ the wall is always at $z=0$ (Fig.~\ref{fig:schematics}(a)),  where $\theta=\theta_e(t)$ (Fig.~\ref{fig:schematics}(c)). In order to calculate the height of the fluid film $h(z, t)\,,$ Eq.~\eqref{eq:pressure} needs two boundary conditions: (i) $\partial h/\partial z\rightarrow \infty$ at $z=L(t)\,,$ (ii) $h(z=0, t)=h_i(t)$ in the moving frame, where $h_i(t)$ is the time dependent height of the solid-liquid interface (Fig.~\ref{fig:schematics}(d)). In the rest frame, $h_i$ satisfies $\partial h_i/\partial t=0\,,$ since the deposit is assumed to be incompressible. In dimensional units, this condition is rewritten in the moving frame as
\begin{equation}
\label{eq:theory14a}
\frac{\partial h}{\partial t}=C \frac{\partial h}{\partial z}\bigg|_{z=0}\,. 
\end{equation}
Eq.~\eqref{eq:theory14a} satisfies the local conservation of the solute mass when the particle concentration at the interface is a slowly varying function in space and time. 

We describe our model in dimensionless units for convenience, indicated using tildes, by $z\equiv \ell\tilde{z}$, $h\equiv H \tilde{h}$, $L\equiv\ell\tilde{L}$, and $t\equiv\tau\tilde{t}$ (see Table II for the length and time scales).  Using the boundary condition $\partial \tilde{h}/\partial \tilde{z}\rightarrow \infty$ at $\tilde{z}=\tilde{L}(\tilde{t})$ (equivalent to $\theta\rightarrow \pi/2$), dropping the tildes and integrating the dimensionless form of Eq.~\eqref{eq:pressure} when $p=p_0$ yields the first-order equation
\begin{equation}
\label{eq:theory14}
\sin\theta=1-\left(z-L\right)^2\,,\quad\text{where}\quad\theta=\tan^{-1}{\left(\epsilon\frac{\partial h}{\partial z}\right)}\,.
\end{equation}
Similarly, Eq.~\eqref{eq:rate} in dimensionless units may be written as 
\begin{equation}
\label{eq:theory14b}
\frac{dL}{dt}=\epsilon(1-\beta)\,,
\end{equation}
where $\epsilon\equiv H/\ell$ (Table II), and the dimensionless deposition speed is $\beta\equiv C/E_0\,.$ From Eq.~\eqref{eq:theory14}, the distance between the deposition front at $z=0$ and the liquid level inside the bath is obtained as $L(t)=\sqrt{1-\sin\theta_{e}(t)}\,.$ The time evolution of $L(t)$ is determined by imposing its equilibrium size as the initial condition, which is given by $L_0\equiv L(0)=\sqrt{1-\sin\theta_{e, 0}}$~\cite{deGennes}. In dimensionless units, the boundary condition given in Eq.~\eqref{eq:theory14c} is rewritten in the moving frame as
\begin{equation}
\label{eq:theory14c}
\frac{\partial h}{\partial t}=\epsilon \beta \frac{\partial h}{\partial z}\bigg|_{z=0}\,. 
\end{equation}
 
In the banding regime, Eqs.~\eqref{eq:theory14}--\eqref{eq:theory14c} determine altogether the meniscus touch-down location $z_c\equiv L(t_c)-1$ ($t_c$ is the instant of touch-down), which is followed by the formation of two contact lines with vanishing contact angles.  Unless the substrate is perfectly wetting (when $\theta_{e, 0}=0$), these contact lines are out of equilibrium and so start to move with a velocity $U$; one runs into the porous deposit, while the other moves with a dynamic contact angle $\theta_D$ until it is restored to its equilibrium value $\theta_D=\theta_{e, 0}$. For a dynamic contact line with $\theta_{e, 0}\ll 1$, the velocity is given by~\cite{deGennes, deGennes2}
\begin{equation}
\label{eq:theory15}
U=\frac{U^\ast}{6\xi}\theta_D (\theta_{e, 0}^2-\theta_D^2)\,,
\end{equation}
where $U^\ast\equiv\gamma/\mu$ ($U^\ast\sim70$m/s in water), and $\xi\equiv\log{(\ell /a)}\sim 5$ is a dimensionless constant, with $a\equiv b/\theta_a$ (logarithmic cutoff). Here $b$ is the slip length with a molecular size, and $\theta_a$ is the apparent contact angle, measured at $x=a$~\cite{deGennes, deGennes2} ($a\sim 10^{-5}m$, $b\sim 10^{-10}m$, $\theta_a\sim10^{-5}$). The travel time of the dynamic contact line $T_D$ can be roughly estimated as follows: When $\theta_{e,0}\approx 15^{o}$ and $\theta_D\ll \theta_{e, 0}$, assuming that the distance $d_D$ traveled by the contact line is nearly equal to the spacing between adjacent bands, $T_D \approx d_D/U\approx 100\mu m/(U^\ast\theta_D\theta_{e, 0}^2/6\xi)\sim 0.1$s, where we have assumed $\theta_D\sim 1^o$. Eq.~\eqref{eq:theory15} is evaluated more precisely by assuming a wedge-shaped liquid border such that $h=x\theta_D$ (\textit{i.e.} $\kappa=0$)~\cite{deGennes}. $\theta_D=\theta_a$ is extracted from Eq.~\eqref{eq:theory15} by replacing $U$ with $E_
0$, as the velocity of the contact line at the meniscus touchdown will be equal to the rate of decrease of the liquid level inside the container. Then the distance $d$ and time $T_D$ at which the contact line travels from the instant of break-up to the moment of re-equilibration at $\theta_D=\theta_{e, 0}$ are given by $d_D=\int_{t(\theta_a)}^{t(\theta_{e, 0})} U dt$ and $T_D=\int_{t(\theta_a)}^{t(\theta_{e, 0})}dt$, respectively. Local mass conservation at $x=a$ yields $a \frac{d\theta_D}{dt}-\theta_D U=0$ in the frame of the contact line. Defining $d_D\equiv \ell \tilde{d_D}$, $a\equiv \ell \tilde{a}$, $T_D\equiv\tau\tilde{T}_D\,,$ and dropping the tildes leads to the following dimensionless travel distance and time of the contact line
\begin{equation}
d_D=a\int_{\theta_a}^{\theta_{e, 0}}\frac{d\theta_D}{\theta_D}\,,\quad\text{and}\quad T_D=\frac{a A}{\epsilon}\int_{\theta_a}^{\theta_{e, 0}}\frac{d\theta_D}{U\theta_D}\,,
\label{eq:SI1}
\end{equation}
where $A\equiv 6 \xi E_0/U^\ast\sim5\times 10^{-7}\,.$ Then the spacing between two adjacent bands is given by $d=d_S+d_D$. The length $d_S$ of the static left-over tail of the meniscus forms when the meniscus touches down and leaves behind a small fluid tail (shown by the dashed curve in Fig.~\ref{fig:theory1}(a)). This tail evaporates completely over time and its length  contributes to the band spacing $d$. Substituting the numerical values into Eq.~\eqref{eq:SI1} and switching back to real units, the travel distance and time of the contact line are found as $d_D\sim 100\mu m$ and $T_D\sim 6s\,,$ in good agreement with experiments (Table I). The difference in the precise value of $T_D$ with the rough estimation of $T_D\sim 0.1$s is due to the fact that the contact line spends a lot of time to increase its speed when $\theta_D\sim\theta_a$, as there is a singularity at $\theta_D=0$ in Eq.~\eqref{eq:theory15}~\cite{deGennes, deGennes2}.

Given the ratio of the deposition front speed to the evaporation rate $\beta=C/E_0$, the height of the fluid film $h(z, t)$ until touch-down is calculated by solving Eq.~\eqref{eq:theory14} subject to the boundary condition in Eq.~\eqref{eq:theory14c}, while the dynamics of $L$ is obtained by solving Eq.~\eqref{eq:theory14b}. Once the meniscus touches the substrate, a single band with a width $\Delta d$ has formed (Fig.~\ref{fig:schematics}(e)). From that instant on, the dynamics of the contact line is governed by Eq.~\eqref{eq:theory15} until $\theta_D=\theta_{e, 0}$, resulting in a spacing $d$ between bands. When contact line motion ceases at $\theta_D=\theta_{e, 0}$, one cycle is complete. The resulting shape of the band, the associated structural quantities, and the instantaneous meniscus profile are shown in Fig.~\ref{fig:theory1}(a). Our model also predicts the termination of the band by a sharp front as observed in experiments, since the deposition growth vanishes when the meniscus breaks up.

Our minimal theory of banding is in quantitative agreement with experiments for \textit{e.g.} $\beta\sim0.3$ (Table I). Furthermore, we show the dependence of $T\,,$ $h_{max}\,,$ and $\Delta d$ in Fig.~\ref{fig:theory1}(b)--(d) for the range of $\beta$ between 0 and 1, which diverge when $\beta\rightarrow 1\,.$ To investigate the scaling behavior of these quantities when $\beta\rightarrow 1\,,$ we first integrate Eq.~\eqref{eq:theory14b}, which yields $L(t)$ as
\begin{equation}
\label{eq:SI3}
L(t)=L_0+\epsilon(1-\beta)t\,.
\end{equation}
At $t=t_c\equiv T-T_D$, the critical time of meniscus touchdown, $L(t)$ should always stay finite as a function of $\beta\,.$ This leads to the scaling form of $t_c$
\begin{equation}
\label{eq:SI4}
\lim_{\beta\rightarrow1}t_c\sim (1-\beta)^{-1}\,,
\end{equation}
such that when $\beta\rightarrow 1\,,$ $L(t_c)>L_0$ and is finite. Thus, $T$ also diverges with $(1-\beta)^{-1}\,.$ Similarly, the bandwidth is given by $\Delta d= \epsilon \beta t_c$, which, when $\beta\rightarrow 1$, leads to 
\begin{equation}
\label{eq:SI5}
\lim_{\beta\rightarrow1}\Delta d\sim \epsilon \beta (1-\beta)^{-1}\,.
\end{equation}
In the limit $\beta\rightarrow 1$ we can derive a scaling relation for the maximum band height $h_{max}$ as well. While a single band is forming around $h_{max}$, $\partial h/\partial z\ll 1$. Then evaluating Eq.~\eqref{eq:theory14} in this limit, and substituting the result in Eq.~\eqref{eq:theory14c} at $z=0$ in the moving frame, we obtain the interface condition
\begin{equation}
\label{eq:SI6}
\frac{\partial h}{\partial t}\bigg|_{h=h_{max}}=\beta \theta_e(t)\,,\quad \text{where}\quad\theta_e(t)=1-L^2(t)\,.
\end{equation}
Integrating Eq.~\eqref{eq:SI6} over time, and in the limit $\beta\rightarrow 1$, we obtain 
\begin{equation}
\label{eq:SI7}
h_{max}\sim (1-\beta)^{-1}\,.
\end{equation}
That is, as $\beta\rightarrow 1$, the time instant at which $h_{max}$ forms should scale again with $t\sim(1-\beta)^{-1}$.

Eqs.~\eqref{eq:SI4},~\eqref{eq:SI5}, and~\eqref{eq:SI7} manifest a continuous transition between the formation of uniform solid deposits and periodic bands. However, the spacing between bands $d\,,$ shown in Fig.~\ref{fig:theory1}(e),  depends only weakly on $\beta$, a consequence of the fact  that this is controlled  solely by the left-over fluid tail (the dashed line in Fig.~\ref{fig:theory1}(a)) when the meniscus touches down. 

This minimal model captures the essential features of banding and filming, as well as relevant time and length scales. However this can only be done in terms of the dimensionless deposition rate $\beta=E_0/C\,,$ which is a free parameter. Therefore, a more sophisticated model is required to determine the dependence of the front propagation speed $C = C(\Phi_b , t)$ on the bulk volume fraction $\Phi_b$ and thence $\beta\,.$  This will further allow the determination of the bandwidth $\Delta d$ as a function of $\Phi_b\,,$ as well as the experimentally observed concentration $\Phi_c$ associated with the transition between the two patterns.

\section{Multiphase flow model for banding and filming}

\begin{table}
\begin{tabular}{|c|c|c|}
 \hline
    \footnotesize{Parameter} &  \footnotesize{Definition} & \footnotesize{Magnitude}  \\ 
   \hhline{|=|=|=|}
  \hline
   $\ell\equiv\sqrt{2}\ell_{cap}$ & length scale & $\sim 10^{-3}$m \\ 
   $\ell_{cap}\equiv \sqrt{\gamma/\rho g}$ & capillary length & $\sim 10^{-3}$m \\ 
   $\tau\equiv H/ E_0$ & time scale & $1-10$~s \\ 
   $\epsilon\equiv H/\ell$ & aspect ratio & $10^{-2}$ \\
   $Ca\equiv \mu E_0/\gamma$ & capillary number & $\sim 10^{-8}$ \\
   $ Pe\equiv E_0\ell/D_s$ & P\'{e}clet number & $10^2$ \\
   $\nu\equiv H/\sqrt{k\mu}$ & scaled deposit thickness & $1$ \\
   $\alpha\equiv \epsilon^3/\nu^3 Ca$ & dimensionless no. & $10^{2}$ \\ 
   \hhline{|=|=|=|}
   $\gamma$ & surface tension & 0.1 N/m \\
   $\rho$ & density of water  & $10^3$ kg/m$^3$ \\
   $g$ & gravitational constant & $10$ m/s$^2$ \\
   $H$  & deposit thickness & $10^{-6}-10^{-5}$~m\\ 
   $E_0$  & evaporation rate & $10^{-6}$~m/s \\ 
   $\mu$  & dynamic viscosity& $10^{-3}$ Pa$\cdot$s \\
       & of water &  \\
   $D_s$    & diffusion constant  & $10^{-11}$~m$^2$/s \\
   $k$    & permeability &  $10^{-9}-10^{-7}$ m$^2$/Pa$\cdot$s \\
    \hline
    
\end{tabular}
\caption{{\bf List of parameters.} Length scales, time scales, and dimensionless numbers (above), and auxiliary physical parameters (below).}
\end{table}

\begin{figure}
\centering
\includegraphics[width=1.02\columnwidth, clip=true]{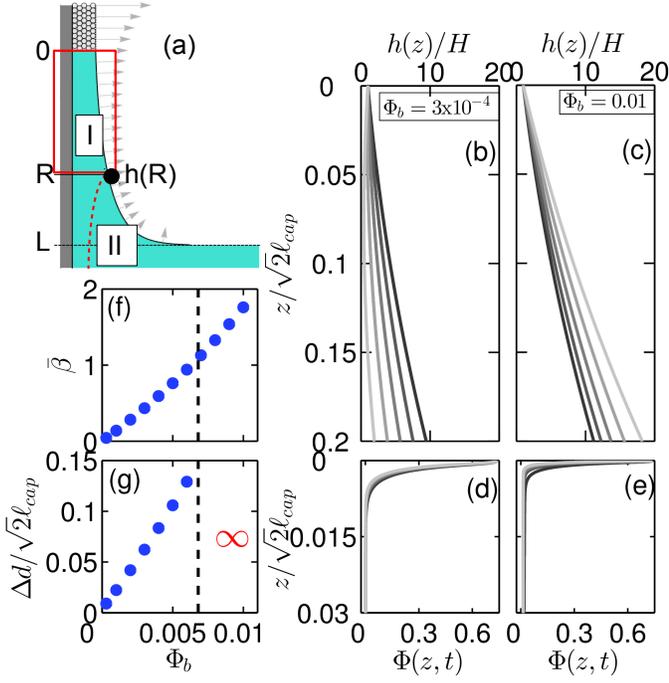}
\caption{{\bf Multiphase model for deposition on a vertical plate in a suspension.} (a) Schematic of the meniscus on the substrate (dark gray). The stagnation point (black dot) lies at the coordinate $(R, h(R))$. The stagnation line (red dashed curve) divides the domain into two flow regimes: (I) capillary-driven viscous shear flow, (II) recirculation flow. The light gray arrows indicate the evaporation profile. (b), (c) The meniscus height $h(z, t)$ in the moving frame of the deposition interface for small and large initial colloidal concentrations, in the domain $z\in \left[0, R\right]\,,$ as shown by the red rectangle in (a). (d), (e) The evolution of the depth-averaged colloidal concentration $\Phi(z, t)$ corresponding to (b) and (c), respectively.  These results follow from Eqs.~\eqref{eq:theory14},~\eqref{eq:theory17a}, \eqref{eq:theory18a}, \eqref{eq:DarcyBrinkman2}, subject to the boundary conditions given by Eqs.~\eqref{eq:BC1atR}--\eqref{eq:BC3atR} at $z=R$ and Eqs.~\eqref{eq:theoryBeta}--\eqref{eq:theoryBC} at $z=0$. In (b)--(e), the grayscale changes from dark to light with increasing time.  (f) shows the mean dimensionless deposition speed $\bar{\beta}$ as a function of the bulk volume fraction $\Phi_b$, where the time average is calculated over an interval $t/\tau\in[4, 14]\,.$ The time when the interface velocity reaches a quasi-steady state is given by $t\sim4\tau\,.$ The points in (f) correspond to $\Phi_b\in\left\{3\times10^{-4}\cup\left[10^{-3}, 0.01\right]\right\}$ increasing in increments of  $10^{-3}$ from bottom to top. (g) shows the bandwidth $\Delta d$ as a function of $\Phi_b\,.$ The points in (g) correspond to $\Phi_b\in\left\{3\times10^{-4}\cup\left[10^{-3}, 6\times 10^{-3}\right]\right\}$ increasing in increments of  $10^{-3}$ from bottom to top. The dashed lines at $\Phi_c=6.8\times10^{-3}$ in (f) and (g) denote the phase boundary between banding and filming, namely when $\bar{\beta}=1\,.$ In (g), the red $\infty$ sign represents the unbounded growth of the bandwidth $\Delta d$ in 
the filming 
regime ($\bar{\beta}>1$).}
\label{fig:twophase}
\end{figure}

\begin{figure}
\centering
\includegraphics[width=0.8\columnwidth, clip=true]{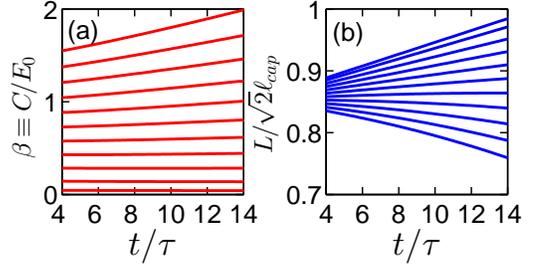}
\caption{{\bf Multiphase model for deposition on a vertical plate in a suspension (continued).} For $\epsilon=0.01$ and $\theta_{e, 0}=15^{o}$ the length of the domain L($t_c$) between the interface and the level of the container at the time of the meniscus break-up $t_c\,.$ These results follow from Eqs.~\eqref{eq:theory14},~\eqref{eq:theory17a}, \eqref{eq:theory18a}, \eqref{eq:DarcyBrinkman2}, subject to the boundary conditions given by Eqs.~\eqref{eq:BC1atR}--\eqref{eq:BC3atR} at $z=R$ and Eqs.~\eqref{eq:theoryBeta}--\eqref{eq:theoryBC} at $z=0\,.$ The time when the interface velocity reaches a quasi-steady state is given by $t\sim4\tau\,.$ The lines in (a) and (b) correspond to $\Phi_b\in\left\{3\times10^{-4}\cup\left[10^{-3}, 0.01\right]\right\}$ increasing in increments of  $10^{-3}$ (a) from bottom to top, (b) from top to bottom.}
\label{fig:twophase2}
\end{figure}

To determine $\beta=C(\Phi_b, t)/E_0$ and the bandwidth $\Delta d$ in terms of the bulk volume fraction $\Phi_b\,,$ as well as the critical concentration $\Phi_c$ at the banding--filming transition, we develop a multiphase flow model of colloids dissolved in a container of liquid. This approach couples the inhomogeneous evaporation at the meniscus and the height of the deposit to the dynamics inside the suspension, \textit{i.e.} the fluid flow and the particle advection. An essential component of this dynamics is the change in the particulate flow from the Stokes regime at low colloidal concentrations away from the deposition front, to a porous flow characterized by the Darcy regime in the vicinity of the deposition front. The porous region is itself created by the particle advection towards the contact line as the suspension turns first to a slurry and eventually a porous plug over the course of drying.

In a meniscus that forms on a vertical plate dipped in a suspension or liquid, there exist two qualitatively different flow regimes when evaporation is present~\cite{Scheid, Stone}. Near the contact line, the thickness of the liquid film is much smaller than its lengthwise dimension.  In this regime (I in Fig.~\ref{fig:twophase}(a)), there is a capillary-driven viscous shear flow which extends partially into the bulk solution, and compensates for the liquid lost by evaporation. In the bulk of the liquid, the meniscus thickness tends to infinity towards the level of the bath. In this region (II in Fig.~\ref{fig:twophase}(a)) there is a recirculation flow where the transverse component of the flow is dominant~\cite{Scheid, Stone}. These two regions are separated by a stagnation curve which terminates at a surface stagnation point $h(R), R <L$ on the liquid air-interface. For small capillary numbers $Ca\ll 1$ (Table~II), we suppose that in region II, namely between $R$ and $L\,,$ the distortion of the fluid-air interface due to the recirculating fluid flow is negligible, resulting in a local hydrostatic profile. This may be justified using the method of matched expansions for $Ca\ll 1$~\cite{Wilson}. Thus, the value $h(R)$ of the meniscus height $h(z, t)$ is determined by hydrostatics at $z=R$ (see Eq.~\eqref{eq:theory14}).

For $h(R) \ll \ell_{cap}$, the viscous forces in the $z-$direction are balanced by the pressure gradients along the slender film in the domain $z\in[0, R]$ (region I), and the forces in the transverse direction are negligible. This simplification of the Navier-Stokes equations is known as the lubrication approximation~\cite{Oron}. In this limit, the problem becomes one dimensional by averaging the local particle volume fraction $\phi(x,z,t)$, local solute velocity $v_{s}(x, z, t)$, and local solvent velocity $v_{f}(x, z, t)$ over the meniscus height $h(z, t)\,.$ Then, the depth-averaged solute volume fraction is given by $\Phi(z, t)\equiv h^{-1}\int^h_0 \phi(x,z,t) dx\,,$ the depth-averaged solvent volume fraction is $1-\Phi(z, t)\,,$ the depth-averaged solute velocity $V_{s}\equiv h^{-1}\int^h_0 v_{s}(x, z, t) dx\,,$ and the depth-averaged liquid velocity $V_{f}\equiv h^{-1}\int^h_0 v_{f}(x, z, t) dx\,.$ Furthermore, we assume that the deposition front speed $C(t)$ only varies temporally, whereas the 
evaporation rate is a function of space alone, \textit{i.e.} $E=E(z)$. In a frame comoving with the deposition front, the depth-averaged equations of local mass conservation for the fluid and solvent are
\begin{equation}
\label{eq:theory17}
\small
\frac{\partial}{\partial t} \left[(1-\Phi) h\right]+\frac{\partial}{\partial z}\left[(1-\Phi) h \left(V_f-C\right)\right]=-E(z)\sqrt{1+(\partial_z h)^2}\,,
\end{equation}
\begin{equation}
\label{eq:theory18}
\small
\frac{\partial}{\partial t} \left[\Phi h\right]+\frac{\partial}{\partial z}\left[\Phi h \left(V_s-C\right)\right]= \frac{\partial}{\partial z}\left[D_s h\frac{\partial \Phi}{\partial z}\right]\,,
\end{equation}
where $D_s$ is the diffusion constant of the solute (Table~II). In Eq.~\eqref{eq:theory17}, we note the presence of an evaporative sink on the right-hand side, with $E(z)= E_0/\sqrt{z/\ell+\Delta d(t)}$ being the singular functional form of the local evaporation rate along the meniscus~\cite{Deegan1, Deegan2}. Here $\Delta d(t)$ is the distance of the wall from the edge of the deposit at a given instant, so that the position of the far edge is given by $z_d=-\Delta d(t)$ in the moving frame. In Eq.~\eqref{eq:theory18}, we note the right side associated with the diffusion of particles (that prevents the formation of an infinitely sharp deposition front); experiments suggest  that diffusion is dominated by advection~\cite{Bodiguel, Jing}, so that the P\'{e}clet number $Pe\equiv E_0\ell/D_s \gg 1$ (Table~II). In addition to the dimensionless quantities introduced in the minimal model of the previous section, we now define the additional dimensionless numbers $\alpha\equiv\epsilon^3/\nu^3 Ca$ and $\nu\equiv H/\sqrt{k\mu}\,,$ where $k$ is the permeability of the porous plug (Table~II). The dimensionless evaporation rate is given by $\tilde{E}(z)\equiv E(z)/E_0$. Then, dropping the tildes, Eqs.~\eqref{eq:theory17} and~\eqref{eq:theory18} in dimensionless form are rewritten as
\begin{equation}
\label{eq:theory17a}
\small
\frac{\partial}{\partial t} \left[(1-\Phi) h\right]+\frac{\partial}{\partial z}\left[(1-\Phi) h \left(\alpha V_f-\epsilon \beta\right)\right]=-E(z)\sqrt{1+\left(\epsilon \partial_z h\right)^2}\,,
\end{equation}
\begin{equation}
\label{eq:theory18a}
\small
\frac{\partial}{\partial t} \left[\Phi h\right]+\frac{\partial}{\partial z}\left[\Phi h \left(\alpha V_s-\epsilon\beta\right)\right]=\frac{\partial}{\partial z}\left[Pe^{-1}\epsilon h\frac{\partial \Phi}{\partial z}\right]\,.
\end{equation}
To complete the formulation of the problem, we need to determine the fluid and particle velocities. In the bulk of the fluid where $\Phi=\Phi_b\ll 1$, the solute and solvent velocities should match ($V_s\approx V_f$), as the particles are advected by the fluid. Beyond the deposition front, the solute velocity $V_s$ must vanish as with $\Phi \rightarrow \Phi_0$, the maximum packing fraction corresponding to the close packing of particles ($\Phi_0\approx 0.74$ for hexagonal packing in three dimensions).  Given that the two limits of Stokes flow and Darcy flow are both linear, to correctly account for both limits and calculate $V_f$ and $V_s$, first we need to determine the depth-dependent velocities $v_f$ and $v_s$. As the particles accumulate near the contact line, the resulting deposit will serve as a porous medium for the fluid. Then, in the lubrication limit, for slender geometries, the transition from the Stokes regime for dilute suspensions ($\Phi\ll 1$) to Darcy flow through porous medium ($\Phi\simeq\Phi_0$) is governed by  the Darcy-Brinkman equation~\cite{Brinkman}
\begin{equation}
\frac{\partial p}{\partial z}=\mu \frac{\partial^2 v_f}{\partial x^2}-\left(v_f-v_s\right)/k\,,
\label{eq:DarcyBrinkman}
\end{equation}
where the pressure $p$ is given by Eq.~\eqref{eq:pressure}.  When the drag term vanishes ($v_s=v_f$), Eq.~\eqref{eq:DarcyBrinkman} reduces to the usual lubrication balance between pressure gradient and the depth-wise shear gradients. In the limit when the particle velocity vanishes and the fluid velocity gradients are dominated by shear against the particles that form a porous plug, we recover the Darcy limit. Since the particle velocity will become vanishingly small as their packing fraction approaches the close-packing limit, this suggests a simple closure  of Eqs.~\eqref{eq:theory17a} and~\eqref{eq:theory18a} $v_s=\left[1-\left(\Phi/\Phi_0\right)^\Gamma\right]v_f$~\cite{Cohen}, which is also valid for the depth-averaged velocities $V_f$ and $V_s$. Here, the exponent $\Gamma$ controls the slope of the crossover between the two regimes.  In combination with the functional relation between $v_f$ and $v_s$, Eq.~\eqref{eq:DarcyBrinkman} can be solved analytically, subject to the stress-free and no-slip boundary conditions $\partial v/\partial x|_{x=h}=0$ and  $v(x=0, t)=0$. Then the depth-averaged speeds are obtained as 
\begin{equation}
\label{eq:theory12}
V_f=\frac{1}{a^3\mu h} \frac{\partial p}{\partial z} \left(\tanh{ah}-a h\right)\,,\quad V_s=\left(1-a^2\mu k\right) V_f\,,
\end{equation}
where $a^2\equiv(\mu k)^{-1}(\Phi/\Phi_0)^\Gamma\,,$ with $1/a$ being the effective pore size. When $a\rightarrow0$ (namely $\Phi\rightarrow0$)  Eq.~\eqref{eq:theory12} reduces to the Stokes expression for the depth-averaged velocity, while when $a \gg 1$ and $\Phi\rightarrow\Phi_0$  we recover the Darcy limit. Defining $V_{s, f}\equiv (\epsilon^2\gamma/\mu \nu^3) \tilde{V}_{s, f}$ and dropping the tildes from the dimensionless velocities $\tilde{V}_{s, f}$, Eq.~\eqref{eq:DarcyBrinkman} becomes
\begin{equation}
\nu^3\frac{\partial p}{\partial z}=\frac{\partial^2 V_f}{\partial x^2}-\left(\frac{\Phi}{\Phi_{\ast}}\right)^\Gamma V_f\,,\quad \Phi_{\ast}\equiv\Phi_0\left(\frac{k \mu}{H^2}\right)^{1/\Gamma}\,,
\label{eq:DarcyBrinkman2}
\end{equation}
in dimensionless units. Here $\Phi_{\ast}$ is a characteristic scaled volume fraction of the colloids, at which the Stokes-Darcy transition occurs; we note that this occurs  before $\Phi=\Phi_0$, the close packing fraction, \textit{i.e.} when the colloids are still mobile~\cite{Cohen}.

The coupled sixth order system of Eq.~\eqref{eq:pressure} in dimensionless form, Eqs.~\eqref{eq:theory17a},~\eqref{eq:theory18a}, and~\eqref{eq:DarcyBrinkman2} constitutes a boundary-value problem and requires the specification of seven boundary conditions in order to find the height of the free surface $h(z,t)\,,$ the particle (and fluid) volume fraction $\Phi(z,t)\,,$ and the deposition rate $\beta$. 
The first boundary condition is given by Eq.~\eqref{eq:theory14} evaluated at $z=R\,,$ 
\begin{equation}
\sin\theta\big|_{z=R}=1-\left(R-L\right)^2\,. 
\label{eq:BC1atR} 
\end{equation}
The second and third boundary conditions at $z=R$ are given by 
\begin{equation}
p=p_0\,,
\label{eq:BC2atR}
\end{equation}
(see Eq.~\eqref{eq:pressure}), and
\begin{equation}
\Phi=\Phi_b\,.
\label{eq:BC3atR}
\end{equation}
Eq.~\eqref{eq:BC1atR} and~\eqref{eq:BC2atR} are the consequence of the liquid-air interface deformations beyond $R$ being hydrostatic. Eq.~\eqref{eq:BC3atR} follows from the fact that $\Phi$ must converge to the bulk volume fraction $\Phi_b$ sufficiently away from the solid-liquid wall. 

At $z=0\,,$  the solute flux should satisfy the continuity condition across the deposition front, which is given by 
\begin{equation}
\label{eq:fluxcontinuity}
\Phi h \left(\alpha V_s-\epsilon \beta\right)-Pe^{-1}\epsilon h \frac{\partial \Phi}{\partial z}=-\epsilon \beta \Phi_0 h\,,
\end{equation}
in the frame moving with speed $\beta\,.$ The left-hand side of Eq.~\eqref{eq:fluxcontinuity} is the flux of colloids at the liquid side of the interface. As the colloids are arrested inside the deposit, the solute flux vanishes as given by the right-hand side. Eq.~\eqref{eq:fluxcontinuity} then yields the deposition rate
\begin{equation}
\label{eq:theoryBeta}
\beta(t)=\frac{1}{\epsilon (\Phi-\Phi_0)}\left(\alpha \Phi V_s -Pe^{-1} \epsilon \frac{\partial \Phi}{\partial z}\right)\,.
\end{equation}
The solvent flowing into the deposit at the interface must replenish the liquid lost due to the evaporation over the solid. In the moving frame this condition in the differential form becomes
\begin{equation}
\label{eq:theory20}
\frac{1}{\epsilon\beta}\frac{\partial}{\partial t}\left[(1-\Phi)h\alpha V_f\right]=- E(z)\sqrt{1+\left(\epsilon\partial_z h\right)^2}\,.
\end{equation}
The remaining boundary conditions at $z=0$ are given by
\begin{equation}
\label{eq:theoryBC}
h=1\,, \quad \Phi=\Phi_i\equiv\Phi_0-2\times10^{-3} \quad\text{where}\quad\Phi_0=0.74\,.
\end{equation}
At the deposit-liquid interface ($z=0$), we set the film thickness constant ($h(0, t)=1$) in dimensionless units. Note that this boundary condition in Eq.~\eqref{eq:theoryBC} is assumed for simplicity, and a fixed deposit thickness as a function of $\Phi_b$ is observed in experiments~\cite{Hatton}.  The deviation of $\Phi_i$ from $\Phi_0$ ensures the asymptotic determination of $\beta$ from Eq.~\eqref{eq:theoryBeta}. Finally, we specify the initial condition of the meniscus height as a hydrostatic profile (see Eq.~\eqref{eq:theory14}) between $z\in\left[0, R\right]$, and assume the initial particle distribution $\Phi(z, 0)=(\Phi_i-\Phi_b)\exp^{-z/z_0}+\Phi_b$ underneath the meniscus, where $z_0\ll 1\,.$ The divergence of the evaporation rate $E(z)= 1/\sqrt{z+\Delta d(t)}$ which is present at $t=0$ is resolved by assuming the initial wall distance $\Delta d(t=0)=10^{-6}\,.$

The numerical values of the dimensionless quantities are given in Table II. We choose $\nu$ to be unity ($\nu=1$) since the effect of bigger $\nu$ on the dynamics is unimportant when $\alpha\gg 1$ holds. We choose the domain size $R=\ell/5$ to ensure that the capillary-driven viscous shear flow regime is dominant. We take the exponent $\Gamma$ as $\Gamma=4\,,$ which models a narrow crossover regime between the Stokes and Darcy flow regimes as a function of $\Phi$.
 
Using the COMSOL finite element package~\cite{Comsol}, we numerically solve Eqs.~\eqref{eq:theory14},~\eqref{eq:theory17a}, \eqref{eq:theory18a}, \eqref{eq:DarcyBrinkman2}, subject to the boundary conditions given by Eqs.~\eqref{eq:BC1atR}--\eqref{eq:BC3atR} at $z=R$ and Eqs.~\eqref{eq:theoryBeta}--\eqref{eq:theoryBC} at $z=0$, for the height of the free surface $h(r, t)\,,$ the particle volume fraction $\Phi(r, t)$, and the deposition front velocity $\beta(t)$. In Fig.~\ref{fig:twophase}(b) we show the time evolution of the meniscus, which corresponds to the formation of a single band with $\Phi_b=3\times10^{-4}$. Here, since $\Phi_b<\Phi_c$, the interface velocity satisfies $\beta<1$ (Fig.~\ref{fig:twophase2}(a)), leading to $\dot{L}>0$ (Fig.~\ref{fig:twophase2}(b)). Therefore, $h(R)$ decreases monotonically, resulting in an overall decrease in the height of the fluid film along $z<R$. Hence, the film surface will approach the substrate over time, followed by the meniscus break-up as exemplified in Fig.~\ref{fig:twophase}(b). For $\Phi_b=3\times10^{-4}$ the break-up location is at $z_c=0.075\approx L(t_c)-1$, in agreement with the global minimum of a quasi-hydrostatic profile.  In  Fig.~\ref{fig:twophase}(c) we show the formation of a continuous deposit for a much larger bulk concentration with $\Phi_b=0.01> \Phi_c$. Here $\beta>1$ (Fig.~\ref{fig:twophase2}(a)), leading to $\dot{L}<0$ ((Fig.~\ref{fig:twophase2}(b))). In this regime a continuous solid layer forms with a constant thickness as dictated by the fixed height condition at the solid-liquid boundary. 

In Figs.~\ref{fig:twophase}(d) and (e), the $z$-dependence of the colloidal volume fraction $\Phi$ is demonstrated inside the meniscus for $\Phi_b=3\times10^{-4}$ and $\Phi_b=0.01$. On both sides of the phase boundary (namely when $\beta=1$), at $\Phi_b=\Phi_c$, $\Phi$ changes rapidly near the solid-liquid interface, which is a natural result of the high P\'{e}clet number $Pe$. This behavior shows qualitative agreement with experiments, where near the interface $\Phi$ of the particles is observed to be much lower than $\Phi_0$. 

The dependence of the mean dimensionless interface speed $\bar{\beta}\equiv t_{c}^{-1} \int_0^{t_c} \beta dt$ on $\Phi_b$ is shown in Fig.~\ref{fig:twophase}(f), where $t_c$ is the time of meniscus touch-down in Fig.~\ref{fig:schematics}(e). The bulk volume fraction $\Phi_b$ at which $\bar{\beta}=1$ corresponds to $\Phi_c$ (Table I). When $h_i$ is constant for all deposition speeds, the bandwidth $\Delta d$ (Fig.~\ref{fig:twophase}(g)) depends linearly on $\Phi_b$ when $\bar{\beta}<1\,,$ and becomes infinite in the filming regime $\bar{\beta}\geq 1\,.$ This behavior implies an abrupt transition in terms of $\Phi_b\,,$ preempting the continuous transition accompanied by the diverging behavior suggested by the minimal model. 

\section{Conclusion}

Our direct observations of the dynamics of the meniscus, contact line, and the shape of the colloidal deposits upon evaporation of dilute colloidal suspensions lead to a simple picture of how deposition patterns arise in these systems. At low $\Phi_b$, meniscus pinning, deformation, touch-down and depinning leads to periodic bands whose spacing is determined by the relative motion of the interface and the evaporation rate, as well as the dynamics of the receding contact line. Meniscus touch-down does not occur at large $\Phi_b$, leading to a continuous colloidal film. A minimal and a detailed quantitative theory capture the transition between banding and filming, the corresponding critical volume fraction, the deposit growth speed, as well as the salient length and time scales, consistent with our observations. Thus, our work reveals the conditions and the dynamics of the concentration-dependent evaporative patterning which has various practical applications~\cite{Tissue1, Hatton, Crosby}.

\begin{acknowledgments}
This research was supported by the Air Force Office of Scientific Research (AFOSR) under Award FA9550-09-1-0669-DOD35CAP and the Kavli Institute for Bionano Science and Technology at Harvard University. 
\end{acknowledgments}
\vskip0.5in

\noindent\textbf{ASSOCIATED CONTENT}
\vskip0.1in

\noindent\textbf{Supporting Information}
\vskip0.1in

\noindent
{\bf Movie 1: Continuous movement of meniscus}

\noindent
The movie shows that the movement of meniscus and particle deposition are continuous when particle concentration is high. Here, one micron latex particles are dispersed in deionized water with a volume concentration of 0.1\%. The field of view is ~470$\times$350 microns. And the movie duration in real time is ~2150 sec.

\vskip0.1in

\noindent
{\bf Movie 2: Periodic formation of colloidal band during solvent evaporation} 

\noindent
A silicon substrate was vertically immersed in a dilute colloidal suspension (PMMA with negatively charged sulfate end groups, 0.002 wt\%, diameter$\sim$ 375 nm). The objective and camera were facing the meniscus and the plane of substrate. As water evaporated naturally, the meniscus moved downwardly in a non-smooth and periodic fashion, leaving periodic bands of colloidal films behind. The whole deposition process shown in the movie took place over $\sim$ 12.5 hours and the visualized screen width is $\sim$4 mm.

\vskip0.1in

\noindent
{\bf Movie 3: Meniscus touch-down and break-up}

\noindent
The formation of a new band on a vertical glass substrate in a suspension via optical microscopy (0.06 vol~\% 1 $\mu$m latex particles in water). The movie shows the meniscus break-up and the subsequent contact line motion. Over time, the interference rings occur, which are associated with the meniscus approaching the substrate. The meniscus break-up starts as a localized event below the maximum of the forming, convex band.



\begin{thebibliography}{}
\bibitem{Opal2} Bones JB, Sanders JV, Segnit ER, Hulliger F (1964) Structure of opal. Nature 204:990-991.
\bibitem{Prum} Prum RO, Torres R (2003) Structural coloration of avian skin: convergent evolution of coherently scattering dermal collagen arrays. J. Exp. Biol. 206:2409-2429.
\bibitem{Photonics1} Blanco A, et al. (2000) Large-scale synthesis of a silicon photonic crystal with a complete three-dimensional bandgap near 1.5 micrometres. Nature 405:437–440.
\bibitem{Photonics2} Rinnie SA, Garcia-Santamaria F, Braun PV (2008) Embedded cavities and waveguides in three-dimensional silicon photonic crystals. Nat Photonics 2:52–56.
\bibitem{Photonics3} Eun Sik K, Wonmok L, Nam-Gyu P, Junkyung K, Hyunjung L (2009) Compact inverse-opal electrode using non-aggregated TiO2 nanoparticles for dye-sensitized solar cells. Adv Funct Mater 19(7):1093–1099.
\bibitem{Tissue1} Sung-Wook C, Jingwei X, Younan X (2009) Chitosan-based inverse opals: Three-dimensional scaffolds with uniform pore structures for cell culture. Adv Mater 21:2997–3001.
\bibitem{Hatton} Hatton B, Mishchenko L, Davis S, Sandhage KH, Aizenberg J (2010) Assembly of large-area, highly ordered, crack-free inverse opal films. PNAS 107:10354-10359.
\bibitem{Snoeijer1} Mar\'{i}n \'{A}G, Gelderblom H, Lohse D, Snoeijer JH (2011) Order-to-disorder transition in ring-shaped colloidal stains. Phys Rev Lett 107:085502.
\bibitem{Bigioni} Bigioni TP, et al. (2006) Kinetically driven self assembly of highly ordered nanoparticle monolayers. Nature Mat 5:265-270.
\bibitem{Deegan1} Deegan RD, et al. (1997) Capillary flow as the cause of ring stains from dried liquid drops. Nature 389:827-829.
\bibitem{Deegan2} Deegan RD, et al. (2000) Contact line deposits in an evaporating drop. Phys Rev E 62:756-765.
\bibitem{Deegan3} Deegan RD, et al. (2000) Contact line deposits in an evaporating drop. Phys Rev E 61:475-485.
\bibitem{Popov} Popov YO (2005) Evaporative deposition patterns: Spatial dimensions of the deposit. Phys Rev E 71:036313.
\bibitem{Crosby} Kim HS, Lee CH, Sudeep PK, Emrick T, Crosby AJ (2010) Nanoparticle stripes, grids, and ribbons produced by flow coating. Adv Mater 22:4600-4604.
\bibitem{Bodiguel} Bodiguel H, Leng J (2010) Imaging the drying of a colloidal suspension. Soft Matter 6:5451-5460.
\bibitem{Adachi} Adachi E, Dimitrov AS, Nagayama K (1995) Stripe patterns formed on a glass surface during droplet evaporation. Langmuir 11:1057-1060.
\bibitem{Stone2} Shmuylovich L, Shen AQ, Stone HA (2002) Surface morphology of drying latex films:multiple ring formation. Langmuir 18:3441-3445.
\bibitem{Chang} Maheshwari S, Zhang L, Zhu Y, Chang HC (2008) Coupling between precipitation and contact-line dynamics: multiring stains and stick-slip motion. Phys Rev Lett 100:044503.
\bibitem{Stone3} Abkarian M, Nunes J, Stone HA (2004) Colloidal crystallization and banding in a cylindrical geometry. J Am Chem Soc 126:5978-5979.
\bibitem{Yodh1} Yunker PJ, Still T, Lohr MA, Yodh AG (2011) Suppression of the coffee-ring effect by shape-dependent capillary interactions. Nature 308:308-311.
\bibitem{Goehring} Li J, Cabane B, Sztucki M, Gummel J, Goehring L (2011) drying dip-coated colloidal films. Langmuir 28:200-208.
\bibitem{Witten} Witten TA (2009) Robust fadeout profile of an evaporation stain. EPL 86:64002.
\bibitem{Kaya} Kaya D, Belyi VA, Muthukumar M (2010) Pattern formation in drying droplets of polyelectrolyte and salt. J Chem Phys 133:114905.
\bibitem{deGennes} de Gennes PG., Brochard-Wyart F., Qu\'{e}r\'{e} D \textit{Capillarity and Wetting Phenomena} (Springer Science+Business Media, New York, NY, USA).
\bibitem{deGennes2} de Gennes PG (1985) Wetting: statics and dynamics. Rev Mod Phys 57:827-863.
\bibitem{Scheid} Scheid B, et al. (2010) The role of surface rheology in liquid film formation. Europhys Lett 90:24002.
\bibitem{Stone} Colosqui CE, Morris JF, Stone HA (2013) Hydrodynamically driven colloidal assembly in dip coating. Phys Rev Lett 110:188302.
\bibitem{Wilson} Wilson SDR (1982) J Engg Math The drag-out problem in film coating theory 16:209-221.
\bibitem{Oron} Oron A, Davis SH, Bankoff SG (1997) Long-scale evolution of thin liquid films. Rev Mod Phys 69:931-980.
\bibitem{Jing} Jing G, Bodiguel H, Doumenc F, Sultan E, Guerrier E (2009) Drying of colloidal suspensions and polymer solutions near the contact line: Deposit thickness at low capillary number. Langmuir 26:2288-2293.
\bibitem{Brinkman} Brinkman HC (1949) A calculation of the viscous force exerted by a flowing fluid on a dense swarm of particles. Appl Sci Res Sect A 1:27-34.
\bibitem{Cohen} Cohen SIA, Mahadevan L (2013) Hydrodynamics of hemostasis in sickle-cell disease. Phys Rev Lett 110:138104.
\bibitem{Comsol} COMSOL 4.3a, Burlington, MA, USA, http://www.comsol.com 

\end{thebibliography}
\end{document}